# Ultranarrow linewidth room-temperature single-photon source from perovskite quantum dot embedded in optical microcavity

Tristan Farrow[1,†,*], Amit R. Dhawan[2,†], Ashley Marshall[1], Alex Ghorbal[1], Wonmin Son[3], Henry J. Snaith[1], Jason M. Smith[2], and Robert A. Taylor[1]

[1]Department of Physics, University of Oxford, Parks Road, Oxford OX1 3PU, UK
[2]Department of Materials, University of Oxford, OX1 3PH, UK
[3]Sogang University, 35 Baekbeom-ro, Mapo-gu, Seoul, South Korea

[†]These authors contributed equally
[*]Correspondence: tristan.farrow@physics.ox.ac.uk

**Ultranarrow bandwidth single-photon sources operating at room-temperature are of vital importance for viable optical quantum technologies at scale, including quantum key distribution, cloud-based quantum information processing networks, and quantum metrology. Here we show a room-temperature ultranarrow bandwidth single-photon source generating polarised photons at a rate of $5\,\mathrm{MHz}$ based on an inorganic $CsPbI_3$ perovskite quantum dot embedded in a tunable open-access optical microcavity. When coupled to an optical cavity mode, the quantum dot room-temperature emission becomes single-mode and the spectrum narrows down to just $\sim 1\,\mathrm{nm}$. The low numerical aperture of the optical cavities enables efficient collection of high-purity single-mode single-photon emission at room-temperature, offering promising performance for photonic and quantum technology applications. We measure 94% pure single-photon emission into a single-mode under pulsed and continuous-wave (CW) excitation.**

**INTRODUCTION**

Ultranarrow linewidth room-temperature (RT) single-photons are essential for photonic quantum technologies[1,2] but their fabrication poses a challenge. Probabilistic single-photon sources such as attenuated lasers are non-ideal[3] while high-performance single-photon emission has only been demonstrated at cryogenic temperatures[4]. Cryogenic cooling is expensive and cumbersome so hinders practical use. Peltier coolers are cryogen-free so offer a cheaper alternative to cryogenic cooling, however, room-temperature operation sets the gold-standard for viable ultranarrow-band single-photon sources. Perovskite quantum dots (PQDs) are promising emitters for cost-effective, scalable, spectrally-pure and colour-tunable single-photon sources for quantum technology applications[5,6]. Quantum confinement in PQDs maintains the non-classical character of the optical signal at RT[7], but like their semiconductor counterparts at higher temperatures their emission linewidth widens by up to tens of nanometres due to phonon-broadening, which undermines their technological potential. Possible strategies have been proposed for producing narrow linewidths (35–65 meV) at room-temperature through targeted chemical treatment of the dot surface to quench low-energy surface phonon modes responsible for broadening[8]. However, restoring the linewidths to almost cryogenic environment-like linewidths at RT is a significantly more demanding task, which may be achieved by constructing a single-photon source comprising an emitter embedded in a tunable optical micro-cavity[9]. This configuration offers the advantages of narrowband emission, excellent emission directionality and high single-mode photon collection at RT. Where light–matter engineering systems such as plasmonic antennas[10–13] demonstrate very high Purcell factor due to ultralow mode volume, open-access optical microcavity systems as demonstrated here offer narrowband single-mode emission and wavelength tunability. Additionally, PQDs can be dispersed in a large range of non-polar solvents after synthesis, so can be spin-coated on a wide variety of surfaces for integration within devices.

We demonstrate such a single-photon source in air at RT featuring an inorganic $CsPbI_3$ PQD (Fig 1a) coupled to an optical micro-cavity (Fig 1c). We observe that the narrowband $TEM_{00}$ mode emission from individual PQDs embedded in the micro-cavity exhibits strong photon antibunching under both continuous-wave (CW) and pulsed excitation with single-photon purity of 94% in a single-mode with ∼1 nm linewidth. In this way we produce bright, pure-colour emission with a detected photon rate of $5 \times 10^6$ per second. We also note the challenges associated with PQDs due to photo-induced degradation or photo-bleaching in intense light fields while chemical passivisation techniques are being developed to improve their robustness[14].

PQD emitters offer excellent optical properties, including fast polarised emission with long coherence times and high quantum yields (95%)[15]. Their near-lifetime limited photoluminescence (PL) linewidth and high quantum yield at cryogenic temperatures offers best-in class performance for an unprocessed nanocrystal system, outperforming by two orders of magnitude typical semiconductor photon sources[16–19]. The emission wavelength of PQDs can be tuned over a wide range (430–730 nm) by modifying their chemical composition, and they maintain optical performance and narrow linewidths up to RT[8]. This, coupled to the low-cost and ease of synthesis, brings them tantalizingly close to industrial scaling-up since most applications operate in air at ambient temperatures.

**RESULTS**

The custom-fabricated open Fabry-Pérot micro-cavities used in this study offer a unique combination of small mode volumes ($< 1\,\mu m^3$) and Q-factors of up to ($> 10^4$)[20], combined with full in situ wavelength tunability of the cavity mode. They consist of a planar mirror, onto which the PQDs are deposited by spin-coating from solution, and a curved mirror, where the distance and angle between the mirrors is controlled using piezoelectric nano-positioning stages. The mirror coatings are tailored to the design-wavelength of the PQDs. The operation wavelengths of the cavities can range from 450 nm to 950 nm and higher depending on the choice of mirror-coating.

**Out-of-cavity measurements**

Photoluminescence from PQD film was characterised using the experimental set-up of Fig 1d) prior to coupling into the optical microcavity. Fig 2*a* and *b* compare the PL peaks of single out-of-cavity PQDs at 4 K in vacuum, and at RT in air. At cryogenic temperatures, their characteristic PL spectrum can be fitted with a Lorentzian profile with linewidth 0.6 nm (1.7 meV), which is within the typical range of 0.6 – 2 meV[21] at cryogenic temperatures for single $CsPbI_3$ nanocrystals with edge length of ∼15 nm. At RT, the out-of-cavity FWHM is more than an order of magnitude wider at ∼40 nm, which is attributed to homogeneous broadening due to low-energy phonon-coupling[8,22] present on the surface of the quantum dots.

Our time-resolved photoluminescence (TRPL) measurements on PQDs (Fig 2d,e) show a typical lifetime of 0.4 ns at 4 K, and 12.2 ns at RT, respectively, which is consistent with observed behaviour[23,24] and is attributed to the fission of excitons into free carriers at higher temperatures[25]. The state lifetime at cryogenic temperature is comparable to the 180–300 ps lifetimes[15,26] reported in lead halide PQDs. We calculated the decay lifetime using a mono-exponential fit of the TRPL curve typical of the transition rate dynamics in two-level systems like PQDs excited at low powers[15,21]. We note that the fast component of the time-resolved PL signal (Fig 2d, e) is more than 2.5 orders of magnitude more intense than the long-lived residual tail of the emission, attributed to delayed carrier recombination during thermalisation and trapping[27]. Detector dark counts account for the flat non-zero intensity segment of the delayed tail of the emission.

The noteworthy optical performance can in part be attributed to the presence of fast, optically-active, triplet states (Fig 1b), present uniquely in lead halide perovskites[21,28,29]. Spin-forbidden triplet transitions delay PL emission, but in these systems they become dipole-allowed due to unusually strong spin-orbit coupling from heavy Pb ions. This results in bright triplet states—the only known example of a material with this property—which can help explain the up-to 1000× brighter PL intensity observed in PQDs compared with other semiconductors. A Rashba-type effect due to symmetry perturbation inverts the energies of singlet and triplet exciton states and lifts the fine structure degeneracy to reveal the ultranarrow linewidths within the fine structure in the orthorhombic and tetragonal phases of the crystal[21,28], but not in the orthogonal phase where the splitting is degenerate. Different PQDs exhibit different decay times, where the variations in lifetime can be attributed to differences in the sizes of the nanocrystals, hence different quantum confinement energies[30].

**Polarised photons**. Polarisation measurements at RT highlight that the PQD emits partially polarised light. Plotting the fluorescence intensity $I(\theta)$ as a function of a linear polariser angle $\theta$ (Fig. 2c) reveals that PQD emission is polarised, which is consistent with other reports[26]. Measured data is fitted to Malus' law, $I(\theta) = I_{min} + (I_{max} - I_{min})\cos^2\theta$, where $I(\theta)$ is the intensity at polariser angle $\theta$, and $I_{max}$ and $I_{min}$ are the maximum and minimum intensities respectively. The degree of linear polarisation, defined as $(I_{max} - I_{min})/(I_{max} + I_{min})$, is found to be 40%. Polarisation of photons in single-photon sources with >50% efficiency and near unity indistinguishability can be achieved with polarised cavities[31]. However, single-photon devices where the source itself is polarised are advantageous in technological applications such as entanglement-based quantum key distribution.

**In-cavity measurements**

**Coupling a PQD to a microcavity**. The planar mirror with spin-coated $CsPbI_3$ PQDs was scanned confocally (Fig. 3a), and individual PQDs, which inherently emit single-photons, were selected for cavity coupling and driven towards the mirror with concave features using nano-positioning motion controllers to create an optical microcavity. This pre-cavity coupling characterization was carried out with the emitter facing the objective to facilitate light extraction. Most PQDs from the tested batch blinked or fluoresced intermittently under 532 nm laser excitation as shown in Fig. 3b. The photo-bleaching of individual PQDs, which is well-known[7], especially in an intense light field such as inside a cavity at RT, can make closing the cavity and recording measurements challenging. The PQDs remain optically active for periods lasting seconds to minutes when the cavity closes due to the increased field intensity and photo-degradation, aggravated by pulsed illumination, after which time the emission becomes too weak for in-cavity measurements. Due to photodegradation of single PQDs, from 100 PQDs, approximately 10 of them could successfully be coupled to the cavity for measurements.

The PQD in the cavity was excited by shining a laser through the planar mirror and the cavity was finely tuned to couple maximum fluorescence from the PQD to the optical cavity $TEM_{00}$ mode. This design featuring a half-symmetric open-access resonator configuration offers two advantages: first, any emitter on the planar mirror can be coupled to a wavelength tunable optical cavity, and second, the concave mirror facilitates optimal coupling by reducing light dissipation due to scattering. Moreover, milling multiple concave features with different radii of curvature on the same plinth permit different coupling possibilities. The cavity-emitted light was collected from the planar mirror side using a 0.85 numerical aperture coverslip-corrected objective. The low angle of cavity emission allows efficient collection even with lower numerical aperture objectives or lenses[32].

The finesse $\mathscr{F}$ of our optical-cavity was recorded to be 100, which yields a quality factor, $Q = q\mathscr{F} = 3 \times 100 = 300$. Here, $q$ is the axial mode index of the optical cavity. Increasing $q$, increases the quality factor and the effective mode volume $V$ (0.5 μm$^3$ in our case), and hinders electromagnetic field confinement in low-width cavities as used here[20,33]. The curved mirror was 4.4 μm wide with a 8 μm radius of curvature. The Purcell factor $F_P = \xi^2 \frac{3\lambda_c^3}{4\pi^2} \frac{Q}{V}$, where $\lambda_c$ is the wavelength of the main cavity mode and $\xi$ is the dipole orientation factor that accounts for the coupling between the emitter and the cavity field. $\xi^2 = 1$ for a perfectly aligned dipole and $\xi^2 = 1/3$ if all possible dipole orientations are averaged. Assuming randomly oriented PQD dipoles, this gives $F_P = 4.7$, and can reach a maximum value of 14 for a perfectly aligned dipole. This moderate value of $F_P$ is attributed to the relatively low $Q$ value of our cavity, which can be increased by using higher finesse cavities.

**Single-mode emission**. The PL spectrum of a single PQD at RT is significantly broader compared to that at cryogenic temperatures. When the PQD is inserted into the microcavity, its emission is forced into the optical modes of the cavity, which acts as a narrow bandpass filter.

By changing the cavity length using piezo-electric actuators to adjust the cavity modes, PQD emission was coupled into a cavity mode, which makes the emission narrowband and single-mode. Fig. 2a displays the emission from a TEM$_{00}$ with an axial mode index of 3. Compared to the 40 nm wide RT free-space emission of a PQD in Fig. 2b, its cavity-coupled emission results in a single-mode with a FWHM of 1 nm. The open-access microcavity design permits straightforward modification of the axial and the lateral emitter position that enables wavelength tunability and coupling of the emitter to different cavity modes. This design has been employed to demonstrate wavelength tunable narrowband RT emission linewidths from other single-emitters as well[32,33]. The coupling of a PQD to a cavity mode leads to a modification of the density of optical states. This Purcell effect alters the spontaneous decay process of the PQD such that its emission is forced into the narrow cavity mode to which it is coupled.

**Single-photon emission**. We performed photon correlation measurements in a Hanbury Brown and Twiss (HBT) setup on the emission from a PQD coupled to an optical cavity TEM$_{00}$ mode (Fig 4a) using CW (Fig 4c) and pulsed (Fig 4d) lasers. Pulsed excitation poses additional challenges for systems prone to photo-bleaching due to the high energy in individual pulses. In both regimes, we recorded 94% single-photon purity. Remarkably, the detected photon rate was 5 MHz—Fig 4b shows the actual photon rate measured by the photon-detector without taking into account any optical system and photodectection losses. Fig 4c shows the time-decay of the PL emission from a single in-cavity PQD in air at RT with a mono-exponential lifetime of 12.7 ns.

## Summary

We have demonstrated a single-photon source in air at RT based on inorganic CsPbI$_3$ perovskite PQDs embedded in a microcavity with a single-photon purity of 94% in CW and pulsed mode operation, generating linearly polarised single-photons at a rate of 5 MHz. Critically, coupling the emission into the cavity mode reduced the emission linewidth to just ~ 1 nm without the need for cryogenic cooling. The reproducible synthesis of PDQs and ease of deposition directly onto cavity surfaces results in a highly reproducible low-cost single-photon system with the potential for transformational impact on quantum technologies at scale with the advent of robust PQDs.

**MATERIALS AND METHODS**

**Mirror fabrication**

An ultra-violet fused silica slip (Spectrosil 2000) was diced to create flat-topped plinth of height 100 $\mu$m, and top area of 300 $\mu$m × 300 $\mu$m on which smooth spherical concave features are created using focused ion-beam milling. Here, we used concave feature of depth 0.3 $\mu$m and radius of curvature of 8 $\mu$m. A planar substrate made of the same material is used for the planar mirror as well. By depositing alternate layers $SiO_2$ and $Ta_2O_5$ by ion-beam sputtering, the dielectric Bragg mirror reflectors are created. The 97.5 ± 0.5% and > 99.9% reflectivity of the planar and plinth mirrors at a central wavelength of 690 nm (selected due to the PQD emission wavelength), respectively allow the creation of an optimum optical microcavity where light is extracted from the planar mirror.

**Quantum dot synthesis**

Reagents: All chemicals were purchased from Sigma Aldrich and used without further purification. Lead iodide ($PbI_2$, 99%), cesium carbonate $Cs_2CO_3$, Reagent Plus 99%), 1-octadecene (ODE, technical grade, 90%), oleic acid (OA, technical grade, 90%), oleylamine (OLAm, technical grade, 70%), methyl acetate (MeAc, anhydrous, 99.5%), octane (anhydrous, 99%) toluene (anhydrous, 99.8%), and ethylenediaminetetraacetic acid (EDTA, ACS Reagent, 99.4%).

Perovskite quantum dots (PQDs) were synthesized following the hot injection method adapted from the literature[34]. Each step up until the PQDs purification was done using standard Schlenk line techniques to keep the reaction air-free, under nitrogen. First, 0.407g $Cs_2CO_3$, 1.25 mL OA, and 20 mL ODE were added to a 100 mL 3-neck flask and degassed for 1 hour under vacuum (flask 1). Flask 1 was then heated to 150°C, the vacuum was switched to an over pressure of nitrogen when the flask temperature was 100°C. Flask 1 was left stirring at 150C until all of the solid $Cs_2CO_3$ was dissolved, indicating that the Cs-oleate had formed. Flask 1 was then cooled to 130°C before being used in the next step.

Into a 250 mL 3-neck round bottom flask, 0.5 g of $PbI_2$ and 25 mL ODE were degassed and then heated to 120°C under vacuum (flask 2). Meanwhile, 2.5 mL of OA and 2.5mL OLAm were heated on a hotplate set at 130°C. The hot OA-OLAm mixture was injected into flask 2 and left under vacuum until all the $PbI_2$ had dissolved. Flask 2 was switched from vacuum to nitrogen and the temperature control unit was set to 180°C. Immediately upon reaching 180°C, 2 mL of the Cs-oleate solution from flask 1 was injected into flask 2. Flask 1 was then moved from the heating mantle to an ice bath as quickly as possible after injection. Once flask 2 had cooled, the reaction was removed from the Schlenk line and exposed to ambient conditions for the purification steps.

The reaction mixture from flask 2 was separated into 2 centrifuge tubes (10 mL in each) and 70 mL MeAc was used to precipitate the PQDs. The PQDs were then centrifuged to form a pellet and the supernatant was discarded. The pellets were redispersed in 5 mL of hexane, then precipitated with 7 mL methyl acetate and centrifuged again. This pellet was dispersed in 2 mL of octane and stored in a glass vial in the fridge. Some precipitate collected on the bottom of the vial overnight, this is avoided when removing the sample from the vial.

**Sample preparation**

For the RT measurements, the $CsPbI_3$ QDs were treated with EDTA by stirring 1 mL of PQDs with 5 mg of EDTA overnight to improve photo-stability and filtered through a 200 nm mesh. Size-selective centrifugation was used in order to obtain monodispersed PQDs. The original PQD solution in octane was diluted by at least 10-fold and then centrifuged at low speeds (2000-3000RPM) for 30 minutes. The resulting supernatant was used for measurements, while the small pellet that formed was discarded. A well-dissolved (in toluene) and concentration-calibrated PMMA solution was prepared and added to the sample which was then spin-coated onto the flat DBR mirror. This resulted in a monodispersed deposition of PQDs at the right concentration (1–5 PQDs/10 $\mu$m$^2$ with a target of 1 PQD/10 $\mu$m$^2$, corresponding to the cavity diameter). Coupling of single-PQDs poses a challenge if the sample is prone to clustering of PQDs, even after calibrating the colloidal concentration to the cavity diameter and laser footprint. Clusters couple to the cavity more readily, revealing its modal structure, but are unsuitable for single-photon generation, which requires the coupling of single PQDs.

A polymethyl methacrylate (PMMA) coat helped to isolate the PQDs from air, albeit with marginal effect as compared to previous attempts where no PMMA was used, and attenuated the signal intensity by approximately 10-15%. The collection efficiency can be improved by centrifuging the PQDs, covering them in PMMA with thickness $\lambda_c$, and by replacing the final SiO$_2$ layer of the planar DBR, since the refractive indices of the two materials match. Chemical passivation such as with Ethylenediaminetetraacetic acid (EDTA) [14] offers an additional strategy for improving the durability of the PQDs against bleaching. We note that there are no restrictions on the operation-temperature of the cavities, since they can work as well in the cryogenic regime as they do at RT.

For the 4 K measurements, solutions of CsPbI$_3$ nanocrystals in toluene were spin-coated at 4000 rpm for 30 seconds onto glass substrates. Various dilutions (in toluene) were trialed until the concentration allowed for the resolution of the emission spectrum from individual PQDs.

**PL, TRPL, and polarisation measurements**

The optical properties of the PQDs at RT were characterized using a confocal micro-photoluminescence setup. The PQDs were excited with a 532 nm Oxiuss diode-pumped solid-state CW laser, and a 532 nm PicoQuant PDL800-D pulsed laser at a repetition rate of 5 MHz with a pulse width of 50 ps. The two microcavity mirrors were mounted on Thorlabs Nanomax 300 piezo-electric stages to control the cavity configuration. Both the laser pump beam was excitation and cavity fluorescence collection was through the planar mirror side using a cover-slip corrected 0.85 numerical aperture Olympus LCPLFLN100XLCD objective. For the out-of-cavity measurements, the same setup was used but without the curved mirror and the emitter, placed on the planar mirror, faced the objective. Single-photon detection and counting was performed using Exelitas SPCM-AQRH-14 SPADs and Swabian Instruments Time Tagger 20.

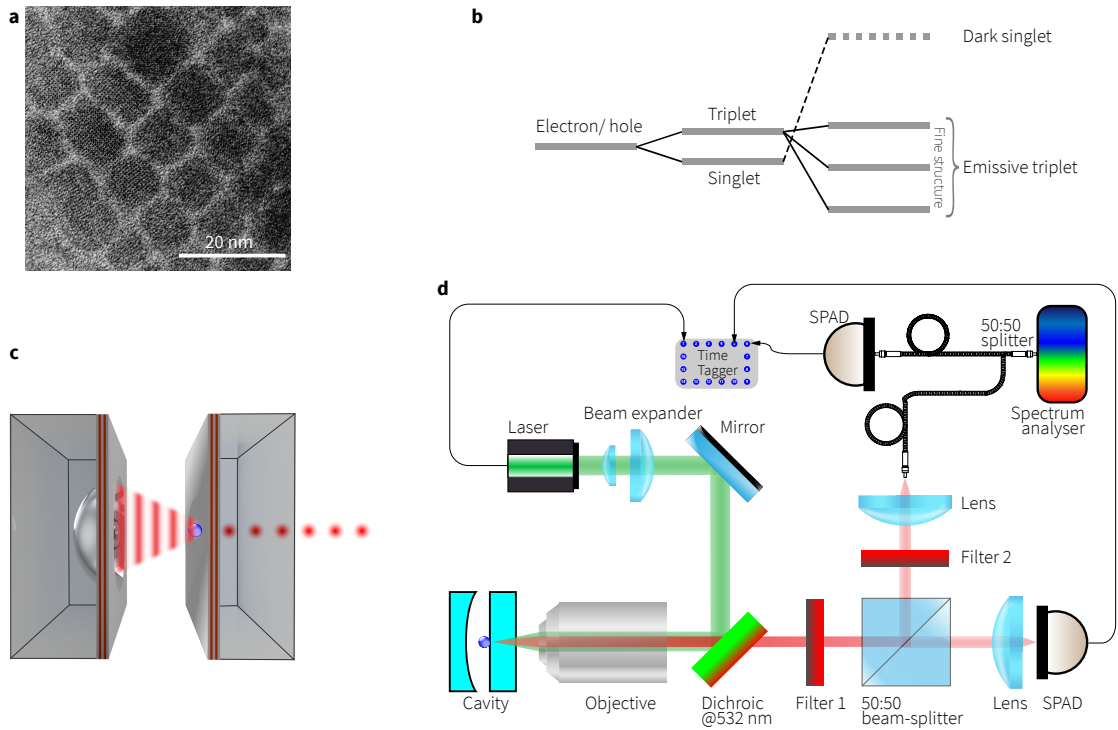

Figure 1: **a**, Transmission electron image of a superlattice of $CsPbI_3$ showing individual PQDs and their size. **b**, Energy level diagram illustrating the exciton fine structure typical of $CsPbX_3$ PQDs with an emissive triplet and a dark singlet on an inverted energy ladder resulting from the interplay between strong spin-orbit coupling and symmetry breaking. **c**, Illustration of PQD–microcavity coupling. The open-access optical microcavity is formed by a planar mirror (supporting the emitter) and a concave mirror. Wavelength tuning is achieved by changing the distance separating the mirrors. **d**, Layout of the optical setup used for cavity coupling and characterization. Filter 1 is a 665±75 nm bandpass filter to allow PL emission to the SPADs (single photon avalanche photodiode), and Filter 2 is a 650±100 nm bandpass filter to prevent optical crosstalk.

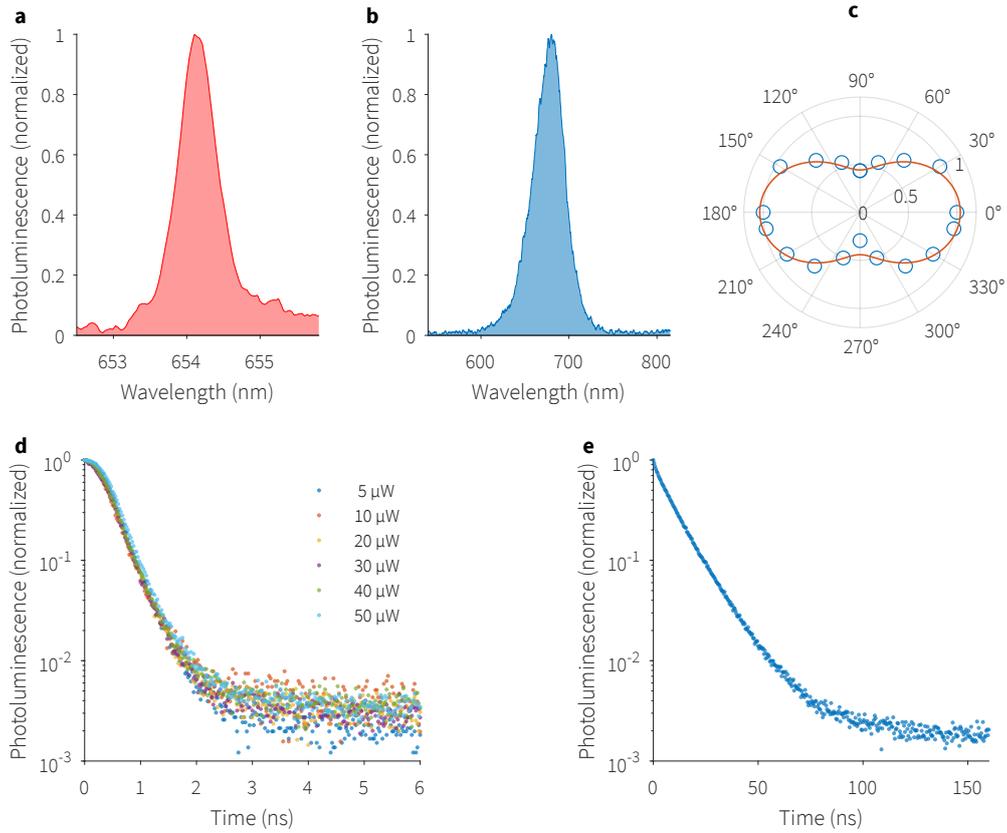

Figure 2: **a**, Out-of-cavity photoluminescence spectrum of a PQD at 4 K showing a FWHM of 0.6 nm. **b**, RT emission spectrum of a PQD, which shows a FWHM of ~40 nm. Polarisation measurement shows a 40% degree of linear polarisation in **c**. Out-of-cavity TRPL measurements on a PQD, **d** at different powers at 4 K, and **e** at RT under 0.3 μW, showing lifetimes of 0.4 ns (typical) and 12.2 ns, resp.

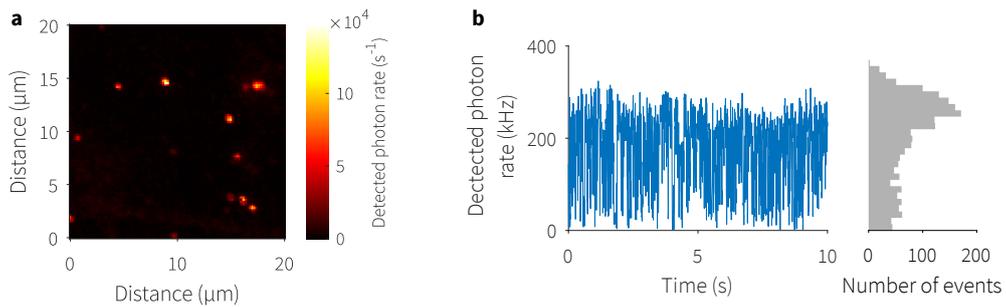

Figure 3: **a**, Out-of-cavity confocal laser scan of a region on the planar mirror spin-coated with PQDs. Well-separated single PQDs are selected for cavity coupling. **b**, typical photoluminescence blinking of a PQD from the used batch — the histogram bin width is 5 ms. The PQD was under 532 nm CW excitation.

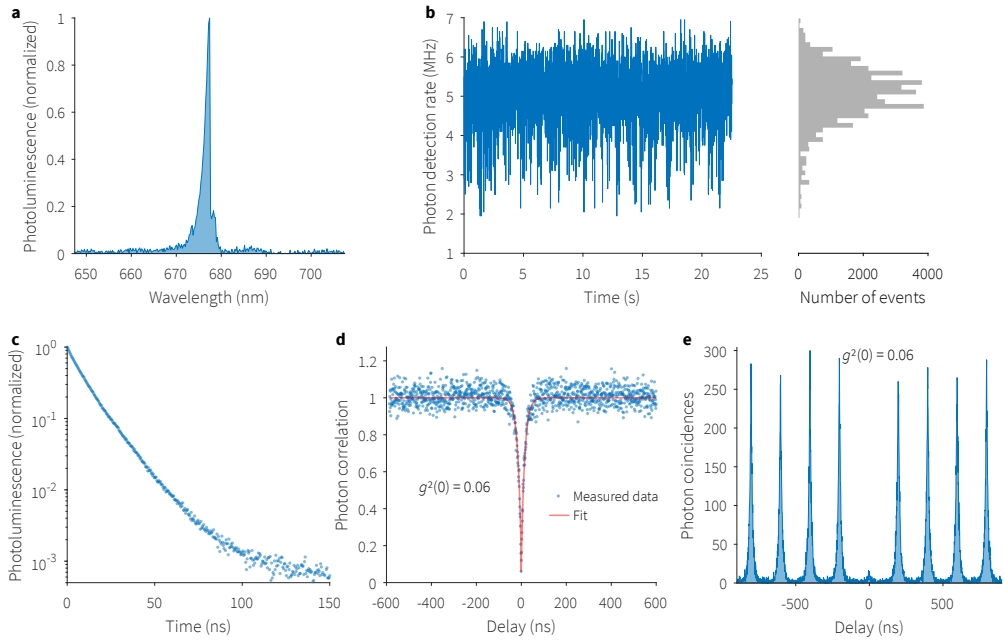

Figure 4: **a**, RT emission from a PQD coupled to a TEM$_{00}$ mode of the optical cavity centered at 677 nm and with a FWHM of 1.1 nm, decayed mono-exponentially with a lifetime of 12.7 ns **c**, and emitted single TEM$_{00}$ mode single-photons with a purity of 94% as shown by continuous **d** and pulsed **e** HBT measurements. **b**, Detected photon rate of 5 MHz (histogram bin width of 5 ms) under CW pumping.